**Structural evolution and kinetics in Cu-Zr Metallic Liquids**

Logan Ward[1], Dan Miracle[2], Wolfgang Windl[1], Oleg Senkov[2], Katharine Flores[3]

[1] Department of Materials Science and Engineering, The Ohio State University, Columbus, OH USA[*]

[2] Materials and Manufacturing Directorate, Wright-Patterson AFB, OH USA

[3] Department of Mechanical Engineering and Materials Science, Washington University, St. Louis, MO USA

Abstract

The atomic structure of the supercooled liquid has often been discussed as a key source of glass formation in metals. The presence of icosahedrally-coordinated clusters and their tendency to form networks have been identified as one possible structural trait leading to glass forming ability in the Cu-Zr binary system. In this work, we show that this theory is insufficient to explain glass formation at all compositions in that binary system. Instead, we propose that the formation of ideally-packed clusters at the expense of atomic arrangements with excess or deficient free volume can explain glass-forming by a similar mechanism. We show that this behavior is reflected in the structural relaxation of a metallic glass during constant pressure cooling and the time evolution of structure at a constant volume. We then demonstrate that this theory is sufficient to explain slowed diffusivity in compositions across the range of Cu-Zr metallic glasses.

1. Introduction

Especially in the past decade, there has been a large amount of discussion as to whether the good glass-forming ability of specific metallic compositions can be directly traced to atomic structure. This idea dates back to a hypothesis by F.C. Frank that the stability of supercooled liquids can be explained by the prevalence of efficiently-packed clusters that have symmetry incompatible with crystal formation, and that icosahedrally-coordinated clusters fill this role for single element systems.[1] More recent studies have considered how these efficiently-packed clusters group together to form medium range order in metallic glasses.[2,3] Along with the energetic stability provided by these clusters, their effect on the atomic mobility in the supercooled liquid has also been studied as a contributing reason for glass formation in metals.[4]

[*] Present Address: Department of Materials Science and Engineering, Northwestern University, Evanston, IL USA

In particular, there has been a focus on the role of icosahedrally-coordinated clusters as a basis for beneficial structural order. These clusters were found to be present in supercooled liquids of multi-component alloys using X-ray diffraction.[5] Via atomistic simulation, it was found that these clusters tend to form large, interconnecting networks in compositions known for forming bulk metallic glasses.[6] That discovery has spurred a focus towards determining whether icosahedral clusters lead to slowed diffusion and therefore bulk metallic glass formation. In particular, it has been shown that the slowest atoms in a Cu-Zr liquid are generally associated with icosahedral clusters.[4] It is also known that these clusters tend to agglomerate and form medium-range order[6], which has been found to further reduce atomic mobility in the liquid.[7] Consequently, it is theorized that the formation of icosahedral short range order and networks can lead to glass formation in at least Cu-Zr-based alloys.

In this work, we take a critical look at this structural explanation of glass-forming ability using the Cu-Zr binary as a trial system. We focus on how structures predicted using molecular dynamics changes with composition, and attempt to draw trends with glass-forming ability. Additionally, we study the evolution of the amorphous structure with temperature during a constant-pressure quench from the liquid. We also perform a rapid quench from the liquid followed by a constant temperature and volume evolution to decouple the effects of temperature and density change from the effect of structural rearrangement. Our overall goal is to assess whether the present theories are sufficient to explain glass-forming ability and structural rearrangements during cooling in this system. We then explore packing efficiency and "defective" clusters as an alternative structural explanation.

2. Methods

Copper-zirconium was chosen as a test system because metallic glasses are known to form between 25 at% to 72 at% copper with several compositions that can be formed in fully-amorphous castings over 1 mm in diameter, including alloys near $Cu_{64}Zr_{36}$, $Cu_{60}Zr_{40}$, $Cu_{56}Zr_{44}$, $Cu_{50}Zr_{50}$, $Cu_{46}Zr_{54}$, and $Cu_{45}Zr_{55}$.[8–11] Also, many of the previous modeling studies discussed in Sec. 1 have been performed for this system[4,6,7], allowing direct comparison of our results. Additionally, interatomic potentials have been specifically developed to model liquid and glassy Cu-Zr, which allows for the use of classical molecular dynamics in this system.[12] For our purposes, using *ab initio* calculations would have severely limited the ability to study systems with large number of atoms on a long timescale.

2.1. Molecular Dynamics

All simulations in this study were performed using LAMMPS.[13] Bonding was described using Finnis-Sinclair type manybody potentials developed by Mendelev *et al* to model liquid and amorphous Cu-Zr.[12] Unless otherwise mentioned, temperature and pressure were controlled using a Nosé-Hoover thermostat and an integration timestep of 2 fs was used.

2.2. Model Generation

Models of metallic glasses were generated by starting with a BCC zirconium supercell containing 11,664 atoms that was randomly seeded with copper atoms to create compositions ranging from $Cu_{40}Zr_{60}$ to $Cu_{80}Zr_{20}$. This wide range was chosen to sample models on both sides of the isostructural composition and, specifically, to include all compositions that are known to form bulk metallic glasses.[8] Compositions were chosen outside of the normal glass forming range (Zr = 20, 30, 60 at%), at compositions known for high glass-forming ability (Zr = 35.7, 44, 50, 54 at%), and compositions near those maxima (Zr = 38.2, 40, 46, 52, 57 at%)[8].

After initial relaxation, each model was heated under constant, zero pressure to 3000 K at $6.75 \times 10^{13}$ K/s, held for 40 ps at that temperature, and then quenched to 1500 K at $7.5 \times 10^{11}$ K/s. At that point, the model was cooled in steps of 32.5 K at a nominal rate of $1.0 \times 10^{11}$ K/s to 200 K. Starting at 1273 K, the time held at each temperature was increased by 12% from the previous step until 623 K. Below that temperature, the hold time was decreased to the pre-1273 K length. This quench schedule was chosen to allocate extra simulation time to the supercooled liquid regime (between approximately 700 K and 1000 K), where most of the structural arrangement occurs. Finally, each model was quenched to 0 K under constant pressure using a conjugate-gradient relaxation technique.

2.3. Structural Analysis

The structure of each model was determined using a radical Voronoi tessellation.[15] The atomic radii for copper and zirconium were taken to be the atomic radii of the pure metals, 1.26 and 1.58 Å.[8] After using these radii to create a radical Voronoi tessellation model, faces smaller than 1% of the total surface area of each cell were removed in order to compensate for overcounting due to numerical issues inherent in the tessellation.[16] Atoms were then defined as nearest neighbors if they shared a face on the tessellation. Using the resultant map of neighbors for each atom, the clusters surrounding each atom were identified and categorized using the Voronoi index, $\langle n_3, n_4, n_5, n_6 \rangle$, where $n_x$ is the number of atoms in the nearest-neighbor shell of the cluster that are bonded to $x$ other atoms in that shell.[17] To characterize network formation, clusters are defined as connected if any atoms of the two clusters are nearest neighbors. This is consistent with medium range order as described in a topological model for metallic glasses that views glass structure as an aggregation of independent clusters[2], but not with other atomistic studies where only clusters that share atoms are defined as connected[3]. Once connections between clusters are determined, it is possible to locate distinct networks within the atomic structure.

### 2.4. Structural Evolution at Constant Temperature

In order to partly decouple the effects of temperature and structural evolution during cooling, a rapid quench from the superheated liquid at 1500 K to the supercooled liquid at 800 K was performed at $3.5 \times 10^{15}$ K/s under constant pressure and controlled temperature for cell compositions of $Cu_{64.3}Zr_{35.7}$ and $Cu_{50}Zr_{50}$. These compositions were chosen because they both are known for high glass-forming ability, but have a large enough difference in composition that they should have much different structures. The final hold temperature at 800 K is slightly above the glass transition for both compositions, which were determined from the volume-temperature behavior to be 765 K and 710 K, respectively. Constant volume was maintained during the structural evolution at the final hold temperature, and snapshots of the atomic positions were saved every 120 timesteps (480 fs).

### 2.5. Diffusivity Calculation

During the simulations of structural evolution at constant temperature described in Section 2.4, parallel simulations were started every 60,000 timesteps (240 ps) in order calculate the mean squared displacement rate of atoms at different points during the structural evolution. In these calculations, the structure was evolved using the same ensemble (constant volume, constant temperature at 800 K) and snapshots of the atomic position were saved every 3 ps for 150 ps. Diffusivity was calculated at several points during the hold by measuring the mean squared displacement of atoms of each type, and using the relationship

$$D = \lim_{t \to 150 \text{ps}} \frac{1}{6} \frac{\delta}{\delta t} \left\langle \left| r(t+t_0) - r(t_0) \right|^2 \right\rangle. \tag{1}$$

### 2.6. Atomic Packing Efficiency

Atomic Packing Efficiency (APE) is a parameter used to approximate how much the packing efficiency of a cluster deviates from ideal.[18,19] Unlike a direct calculation of volume packing efficiency, this parameter allows assessment of how close the system is to optimal packing, independent of cluster configuration and strain applied to the cluster. APE is defined as the radius ratio between the central atom and the average radius of atoms in the nearest-neighbor shell, normalized by the ideal radius ratio for a cluster with that number of atoms, which has been established in literature.[20] In this notation, the parameter is defined as

$$APE = \left(\frac{r_0}{r_1}\right) \bigg/ R^* \qquad (2)$$

where $r_0$ is the radius of the central atom, $r_1$ is the mean radius of the nearest neighbors, and $R^*$ is the optimal radius ratio of that cluster.[8,18] An ideally-packed cluster has an APE of 1, a cluster with excess free volume has APE > 1, and one packed too closely has APE < 1.

2.7. Cluster Volume Packing Efficiency

The packing efficiency of a cluster is defined as the volume attributed to atoms inside a cluster divided by the total volume of the cluster, and was calculated in a similar manner to previous work by Yang *et al.*[21] The total volume can be determined by finding the volume of the convex polyhedron whose vertices are given by the centers of atoms in the nearest neighbor shell. In order to determine the volume of the atoms inside the clusters, the convex hull polyhedron is first segmented into tetrahedra formed by the atoms at the corners of each face and the central atom. The solid angle of each atom inside each tetrahedron is then calculated using the relation provided by van Oosterom and Strackee.[22] The fraction of the volume of an atom inside of the cluster is equivalent to its total solid angle included in all of the tetrahedra divided by 4π.

2.8. Cluster Lifetime Measurement

To assess the lifetime of a cluster in the liquid, a model of a $Cu_{50}Zr_{50}$ liquid at 800 K was run at constant temperature and volume starting after the quench and hold described in Section 2.4 . The simulation was run for 20,475 timesteps (81.9 ps) with snapshots output every 5 timesteps (20 fs). The atom trajectory was filtered such that any movement attributed to oscillations faster than 5 ps were removed. This was accomplished by calculating the discrete Fourier transform of the x, y, and z components of the trajectory, setting any frequency component higher than 200 GHz to zero, and then calculating the inverse Fourier transform.[†] This filtering was applied to remove thermal vibrations from the measurement of structural rearrangement lifetimes, measured as the contiguous time span during which an atom has the same set of nearest neighbors.

3. Results/Discussion

   3.1. Composition Dependence of Icosahedral Networks

The composition range chosen for our study samples structures with a wide distribution of short range order, as shown by the fractions of the majority cluster types in Figure 1. For example, in the composition range of 20 to 40 at% Zr there is a strong majority of Cu-centered full icosahedra ( $\langle 0,0,12,0 \rangle$ ) and "icosahedron-like" clusters (i.e. $\langle 0,2,10,0 \rangle$ ), while fewer than 4% of clusters have one of these topologies in glasses with higher zirconium content. Taking only the fraction of a single cluster type into account, there is no single cluster that can easily account for the glass-forming ability of copper-zirconium metallic glasses. Full icosahedra, which have been hypothesized to lead to improved glass-forming ability in metals, reach their maximum concentration near $Cu_{70}Zr_{30}$ – a composition that is not known to form bulk metallic glasses. Also, icosahedra are only present in small concentrations near the high zirconium fraction bulk metallic glass compositions.[11] Even considering the other top clusters, no single cluster topology shows a correlation with the compositions known for maximum glass-forming ability, which lie between $Cu_{64}Zr_{36}$ and $Cu_{46}Zr_{55}$.[9–11] At most, the Cu-centered $\langle 0,5,6,0 \rangle$ clusters reach maxima in this region, which could allude to but does not prove their importance in glass-formation.

---

[†] The software used to perform this filtering is available on "http://atomistics.osu.edu/"

This raises the question of whether the coordination of several efficient clusters into a larger structure can explain local regions of improved glass-forming ability. A number of studies have shown that icosahedrally-coordinated atoms tend to agglomerate into networks in some Cu-Zr based metallic glasses.[6,23] These structures have been hypothesized to be the source of slowed dynamics in metallic glasses, as atoms associated with icosahedra tend to diffuse less.[4] While atomic mobility is even slower for agglomerated clusters[7], it has yet to be shown whether agglomerated clusters are unique to bulk metallic glasses or just an intrinsic property of supercooled liquids.

Figure 2 shows the number and average size of networks of linked icosahedral clusters for several Cu-Zr compositions at 0 K. All icosahedra in compositions less than approximately 44 at% zirconium are part of a single network. This large degree of interconnectivity is unsurprising due to the large fraction of icosahedral clusters, which is close to the point (7.6 at%) where it becomes impossible to pack icosahedra without them sharing neighbors. While, based on current literature[6,7,23], these pervasive networks of icosahedra should be key features in metallic glasses, our work shows that they are neither sufficient nor necessary conditions for bulk metallic glass formation. As found with the fraction of icosahedral clusters, the maximum network size occurs outside of the known glass-forming region, near $Cu_{70}Zr_{30}$. There is also an order of magnitude decrease in the average network size between $Cu_{64.3}Zr_{35.7}$ and $Cu_{50}Zr_{50}$, yet both are known to have superb glass-forming ability.[8] This evidence stands in contrast to the current view [6,7,23] that glass-forming ability can be simply attributed to the presence of networks of interconnected icosahedral clusters, and clearly shows that glass-formation does not require these networks and cannot be adequately measured by their size.

While pervasive networks of icosahedral clusters were found not to be unique to bulk metallic glass forming liquids, there is still evidence to support that they play a role in bulk metallic glass formation. Several studies have attributed icosahedral clusters and their networks to locally stunted atomic mobility, but this leaves several questions open. Do these networks affect the *global* diffusivity? Why do these structures form?

3.2. Constant Temperature Evolution

In order to help answer these questions, we developed a technique to study the development of structure at constant temperature (as described in section 2.4), which enables us to decouple the effects of structural rearrangement from changes in temperature or density. Before further testing the importance of network formation with this technique, we applied it to the model of a $Cu_{64.3}Zr_{35.7}$ glass in order to understand how well it replicated the features of a slower quench rate and thereby assess whether the structures are physically reasonable. As a result of the exceedingly high quench rate of over $10^{15}$ K/s before the constant temperature holds, there is little structural change between the initial model at 1500 K and the model at 800 K after the quench. The partial coordination numbers of the Cu and Zr atoms do not change by more than 0.6%, which suggests little to no structural rearrangement took place.

The top five cluster types in the model also do not change before and after the quench, though their order does and the fractions do change by less than 25%. This is fractional change is notable but is very small in comparison to change in cluster fractions with temperature found at a slower quench rate, as shown in Figure 3a. For example, the fraction of icosahedral and quasi-icosahedral clusters increase by approximately 1700% and 400% during a slower quench, respectively. In comparison, they only change by 44% and 66% during the faster quench. After evolving the rapidly quenched model for 2 ns, the fractions of different cluster configurations (see Figure 4) resemble the structure produced during the slower quench and stabilize after approximately 1 ns. Also considering that the density of this liquid is within 2% of the slower-quenched structure, we conclude that the rapid quench freezes in the high-temperature liquid before the start of the hold and still results in a similar structure to a slower quench within accessible simulation times after holding at the lower temperature.

### 3.3. Network Development

In order to study how networks form in metallic glasses, the structures of a $Cu_{64.3}Zr_{35.7}$ metallic glass generated during the simulated quench from 1500 K to 200 K were studied. As shown in Figure 3b, the number of icosahedral networks generally increases from around 30 at 1500 K to around 38 at 1050 K, below which the number of networks quickly approaches one. The average size of the networks, on the other hand, increases continuously with decreasing temperature. The trend is modest from 1500 K to about 1050 K, below which the size increases rapidly. This behavior suggests that the networks grow in number and size until they impinge to form a single network. This hypothesis was tested by studying the constant temperature evolution of the same $Cu_{64.3}Zr_{35.7}$ model as described in Sections 2.4 and 3.2.

As mentioned in Section 3.2, the short range order, viewed as the distribution of different cluster polyhedron topologies, rapidly changes during relaxation at 800 K. As shown in Figure 4, the fraction of icosahedral clusters quickly increases from nearly non-existent to the second most prevalent structural feature in the first 230 ps, which makes the structure similar to that at 800 K for a slower quench rate (see Figure 3a). This fast rate of growth contributes to the quick formation of networks of icosahedral clusters, as shown in Figure 5. The large number of small networks originally present at high temperature drops to only a single network within about 500 ps of the relaxation time. By monitoring which atoms are present in the network, it was determined that the largest network at 500 ps continued to be the largest network at longer time periods and absorbed smaller networks, supporting the hypothesis of network coalescence.

This rapid formation of a single, pervasive network can be attributed to two factors: the large number of icosahedral clusters and the enhanced stability of agglomerated icosahedral clusters. Only 1.5 ns after the quench, nearly 40% of the atoms in the simulation are part of an icosahedral cluster. Additional clusters that form will likely be in close proximity or part of the network even without considering any bias to form as part of a network. Additionally, the network formation could also be enhanced by the increased stability of networked icosahedral clusters, as demonstrated by Hao *et al.*[7] Furthermore, within the time frame of 1500 to 2000 ps at 800 K, the average number of connected icosahedral clusters that remained stable, formed, or dissolved between snapshots of atomic positions 20 ps apart, was determined to be 11.48, 9.43, and 9.41, respectively. This provides evidence that stable icosahedral clusters are better connected, further suggesting the tendency of icosahedral clusters to aggregate into large network.

3.4. Effect of Structural Relaxation at 800 K on the System Energy and Element Diffusivity

Evolving the metallic glass structure at a constant temperature provides the unique ability to measure changes in system energy and global diffusivity as a result of structural change. Since the volume of the system is held constant, it is possible to decouple the effect of structural rearrangement from changes in free volume. As shown in Figure 6a, the total internal energy of the structure rapidly decreases during the first 100 ps and then continues decreasing, but with slower rate. This shows that the structure is relaxing into lower energy states with the system volume held constant. The decrease in internal energy during evolution of the 1500 K structure at 800 K reinforces the idea that the structure in the liquid is strongly dependent on temperature and provides the additional observation that the most efficiently-packed structure should not be viewed simply as the densest on a global scale.

Additionally, the diffusivity of the alloying elements in the amorphous structure was studied as a function of time to assess whether the evolution in the structure can be attributed to stifled diffusion.

As the diffusivity calculation utilized in this study requires analyzing data from 150 ps worth of snapshots, it is not possible to make instantaneous measurements of diffusivity changes. Even so, it was possible to resolve that the diffusivity in the model decreases by 60% within the first 250 ps of the simulation time (as shown in Figure 6b), which is coincident with the growth of icosahedral short range order and networks. Combined with the knowledge that icosahedral clusters locally reduce atomic mobility and this behavior is enhanced for clusters present in networks[4,7], this observation supports the idea that the development of icosahedral order slows diffusivity.

To determine whether this is indeed a pure effect of the icosahedral network formation, the same quench and hold experiment was performed with $Cu_{50}Zr_{50}$ – a composition with similar glass-forming ability to $Cu_{64.3}Zr_{35.7}$ but without the tendency to form large icosahedral networks. As shown in Figure 7a and 7b, the structural change in $Cu_{50}Zr_{50}$ is not as significant in terms of cluster topology or networks of icosahedral clusters in comparison to the copper-rich composition. Icosahedral clusters are far from the most prevalent structural feature; and the network order only increases slightly after quenching. However, there is still a change in the energy of the system over time and a drop in diffusivity after the quench (see Figure 7c and 7d). These small structural changes were found to correlate with a slightly smaller change of 46% in diffusivity 200 ns after the rapid quench in comparison to 56% in $Cu_{64.3}Zr_{35.7}$. This change in diffusivity with only minor change in icosahedral short range order and networks does not disprove that networks of interconnected icosahedral clusters could contribute to slowing diffusion in supercooled liquids, but shows they are not the only reason. In particular, the common opinion that icosahedral clusters are the cause of stifled diffusivity lacks the ability to explain the change in diffusivity and energy of the clusters over time in $Cu_{50}Zr_{50}$.

### 3.5. Alternative Explanation: Efficiently Packed Clusters and Structural Defects

Another way of conceptualizing this structural relaxation is to extend the view of structural defects developed by Egami *et al.*[24] This theory defines 'defective' structural sites as atoms that experience either positive (compressive) or negative (tensile) local atomic stresses and corresponding volume strains. We extend this view here by noting that a defective structural site and its first shell of neighbors produces atomic clusters whose packing efficiencies are either higher (for positive local stresses) or lower (for negative local stresses) than ideal. We can thus track the presence of these defective sites using local packing efficiency and internal energy as proxies for volumetric strains.

The first parameter we use to investigate this concept in our models is the standard deviation of cluster volume packing efficiencies (see Section 2.7). Presumably, there is an ideal packing efficiency for each species in a metallic glass and, based on this theory, the average deviation from that ideal should decrease as the model becomes more relaxed. As shown in Figure 8, the standard deviation of cluster packing efficiency decreases with temperature during quenching from 1500 K to 300 K, which is consistent with the idea that 'defective' sites are eliminated as the glass relaxes at lower temperatures. Additionally, as shown in Figure 9, the standard deviation of the packing efficiency of Cu-centered clusters rapidly decreases with time during the constant volume and temperature holds for both $Cu_{64.3}Zr_{35.7}$ and $Cu_{50}Zr_{50}$. This further provides evidence that we can attribute structural relaxation in the supercooled liquid to the elimination of defective clusters.

In order to further examine the idea that clusters with far-from-ideal packing efficiencies dissipate during relaxation, we analyzed the model structures using the APE parameter.[18] As discussed in Section 2.6, this parameter describes the deviation of a cluster from the ideally packed one where any perfectly-packed cluster has an APE of 1, clusters with excess free volume have APE>1, and those which are packed too tightly have an APE<1. This allows us to determine whether a cluster is close to ideal packing irrespective of the magnitude of that cluster's optimal packing efficiency. In effect, we can then determine if the system as a whole has reached a local maximum packing efficiency even if that consists of multiple cluster types, each with a unique optimal efficiency. This is especially important for amorphous materials, which are inherently structurally heterogeneous.

The mean and standard deviations of the APE of each cluster type both decrease as a function of temperature during a quench from 1500 K to 300 K, as shown in Figure 10. The mean decreases from approximately 1.022 to 1.005 during the quench, which implies that the average cluster loses free volume without becoming "over-packed" during the quench. The idea that a liquid loses free volume during cooling and still has some residual free volume is expected for an amorphous material. The deviation in APE also decreases, which implies a similar conclusion to what was developed based on the volume packing efficiency: elimination of clusters with far-from-ideal packing efficiencies is a key part of relaxation. Additionally, despite the remaining free volume after the relaxation, the magnitude of the standard deviation (~0.06) in combination with the average APE of 1.005 indicates that the final structure should not only have pockets of "free volume" (i.e. lower-density areas, APE > 1), but also higher-density areas (APE < 1).

The change in the APE of clusters was also studied during a constant volume and temperature simulation at 800 K in order to further explore its relationship with structural relaxation. During the first 200 ps of the hold, the mean APE for the $Cu_{64.3}Zr_{35.7}$ model dropped from 1.023 to 1.0098, which implies that the system adopted more efficiently-packed local configurations. This change in packing efficiency with fixed volume is observable because the APE parameter only reflects the deviation of the packing from optimally efficient, and does not change under the strain imposed by the constant volume condition since this measure does not explicitly take the positions of the atoms into account. The new atomic arrangements produced by the model are energetically favorable (as observed in the decrease in internal energy in Figures 6 and 7) and, based on having an APE close to unity, would also be efficiently packed under conditions without the constant volume constraint. Additionally, the standard deviation reduces slightly from 0.080 to 0.063 over the first 200 ps. This shows that the system is rearranging into a new configuration that favors clusters with ideal packing by removing more defective ones. The same behavior is seen during the quench and hold simulation with $Cu_{50}Zr_{50}$, where the mean drops from 1.021 to 1.010 and the standard deviation drops from 8.38% to 6.99% during the first 200 ps. This shows that structural evolution by defect elimination takes place even in compositions with vastly different structural characteristics. Recall that the $Cu_{64.3}Zr_{35.7}$ alloy has a large fraction of icosahedral clusters that form a network, which is not the case for the $Cu_{50}Zr_{50}$ alloy.

The lifetime of clusters in the liquid was studied to assess the effect of efficient packing as measured by the APE parameter on diffusivity. Icosahedral clusters were originally reported to lead to slow dynamics because they were found to have a longer lifespan in the supercooled liquid[7,25], which implies that atoms associated with them move less frequently with respect to their neighbors. As shown in Figure 11, the lifespan of clusters was found to increase slightly as the APE of a cluster approaches unity. While this does not conclusively show that these areas lead to higher diffusivity, it does suggest that ideally packed clusters (APE near 1) could perform the role reported earlier for icosahedral clusters.

3.6. Comparison to Existing Theories

The present work shows that the prevalence of icosahedral short range order and networks does not give a compelling correlation with glass forming composition regions. The formation of icosahedral clusters also cannot explain the decrease in diffusivity associated with the structural relaxation found in the $Cu_{50}Zr_{50}$ alloy. Furthermore, recent work has shown icosahedral clusters are not even the most energetically stable cluster in metallic glasses.[26] Taken together, these arguments suggest that there is no unique advantage in energetic stability or slowed dynamics specific to the formation of icosahedral order in metallic glasses.

The suggestion that the demise of the most defective clusters and the formation of more efficiently packed clusters gives structural relaxation and slowed diffusivity is consistent with the prevailing theory for why icosahedral clusters form and what that implies. However, the present work generalizes structure relaxation and slowed diffusivity in composition ranges where icosahedral clusters are not favored to form. Thus, icosahedra do not exert a unique influence on glass-forming ability, and this effect is shared with other cluster types as well. The unifying feature, regardless of the topology, is the cluster packing efficiency. In this view, icosahedra play a dominant role in the stability of Cu-Zr glasses with ≤ 45 at.% Zr because they are the most efficiently packed cluster in this composition range. An icosahedrally-coordinated cluster is efficiently packed when the ratio between the central atom's radius and the effective radius of the neighbors is approximately 0.902.[20] Copper-centered icosahedral clusters that contain approximately 6 to 8 copper neighbors have the most nearly optimal packing efficiency, with radius ratios of 0.887, 0.904, and 0.922, respectively. These clusters have compositions between 30.8 and 46.2 at% Zr, which is near the composition region where the fraction of icosahedral clusters is the greatest. The maximum fraction of icosahedral clusters near this range could be attributed to a combination of driving forces of the system to minimize energy by increasing packing efficiency and the easy formation of clusters of these compositions given the system concentration. The high fraction of these clusters coupled with the enhanced stability of forming networks explains the formation of networks in that composition range.

In the same way, the stability of Cu-Zr glasses with between 45 and 60 at.% Zr (including $Cu_{50}Zr_{50}$) will be most strongly influenced by the most efficiently packed clusters in that compositional range. As shown by Figure 12, copper-centered clusters with 11 neighboring atoms, $\langle 0,5,6,0 \rangle$ and $\langle 0,7,4,0 \rangle$, have the closest APE to ideal of the most common clusters in this composition range. Correspondingly, they were found to be the most prevalent structural feature in that set over most of the range (see Figure 1).

Additionally, as the liquid approaches a composition of 60 at.% Zr, Cu-centered clusters with 10 neighbors ($\langle 0,6,4,0 \rangle$) and Zr-centered clusters with 14 neighbors ($\langle 0,6,8,0 \rangle$) are the most efficiently packed and most prevalent. This agreement shows that the theory described in this work provides an improved framework for understanding structures in metallic glasses and how they influence the ability of the liquids to form glasses.

4. Conclusion

In this work, we studied the evolution of structure and properties of a Cu-Zr metallic liquid during cooling and evolution at constant volume and temperature. It was found that the development of icosahedral order fails to explain the observed diffusivity and energy changes in $Cu_{50}Zr_{50}$. The present work develops a more general view that ideally-packed atomic clusters are formed at the expense of clusters with excess or deficient free-volume during structural relaxation of a metallic glass, which supports earlier work by Egami *et al.*[24] The proposed model also explains slowed diffusivity in compositions and alloy systems that do not show a propensity for forming icosahedrally-coordinated clusters and provides a framework for understanding why the formation of networks of icosahedral clusters is correlated with slowed dynamics.

5. Acknowledgements

The work at Ohio State and Washington University was funded by the Defense Threat Reduction Agency under Grant No. HDTRA1-11-0047 and by the Air Force Office of Scientific Research under Grant No. FA9550-09-1-0251. Simulations were performed at the Ohio Supercomputer Center (Grant No PAS0072). LW was supported by the Department of Defense (DoD) through the National Defense Science & Engineering Graduate Fellowship (NDSEG) Program.

The work at the Air Force Research Laboratory was supported through the Air Force Office of Scientific Research (M. Berman, Program Manager, Grant Number 10RX14COR) and the Air Force under on-site contract No. FA8650-10-D-5226 conducted through UES, Inc., Dayton, Ohio.

7. Figures

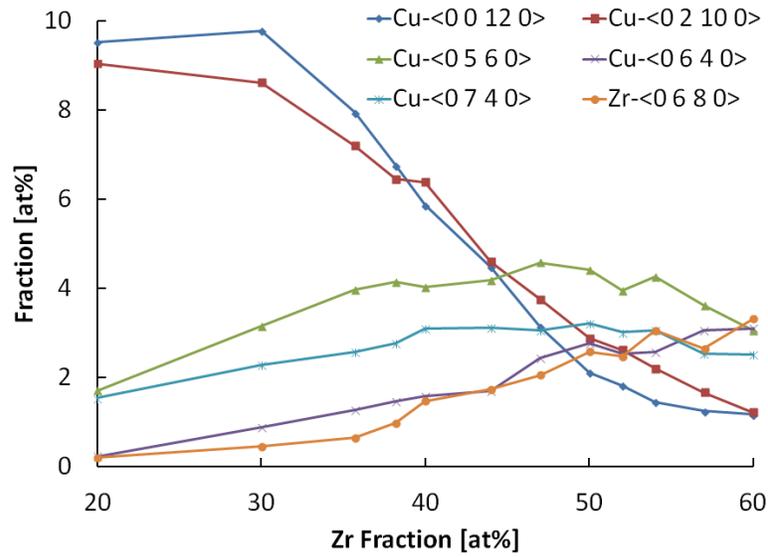

**Figure 1.** The most common atomic clusters identified in the amorphous structures of Cu-Zr metallic glasses with the Zr concentration ranging from 20 to 60 at.%. Each cluster is identified based on the species at the center and the polyhedron index (enclosed in angle brackets). The amorphous structures are the result of a quench from 1500 K to 0 K at approximately $10^{11}$ K/s simulated using classical molecular dynamics.

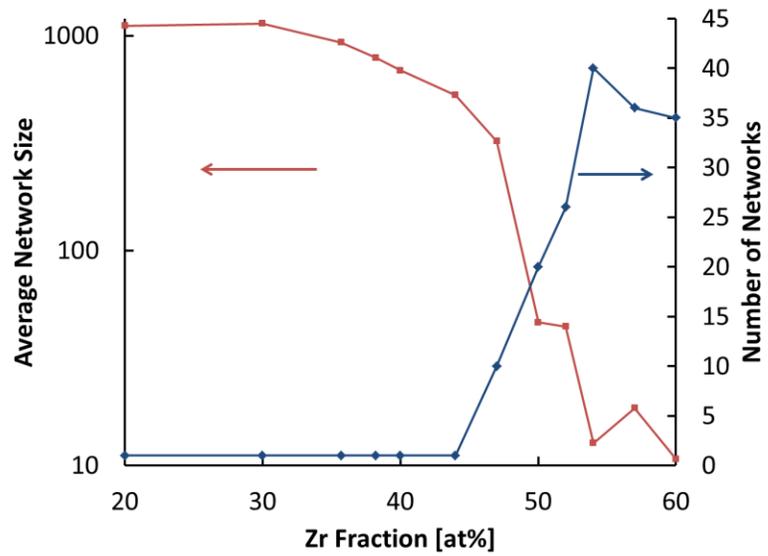

**Figure 2.** Average size and number of networks of icosahedrally-coordinated atoms in Cu-Zr metallic glasses as a function of composition determined from the 0 K models of metallic glasses produced using molecular dynamics.

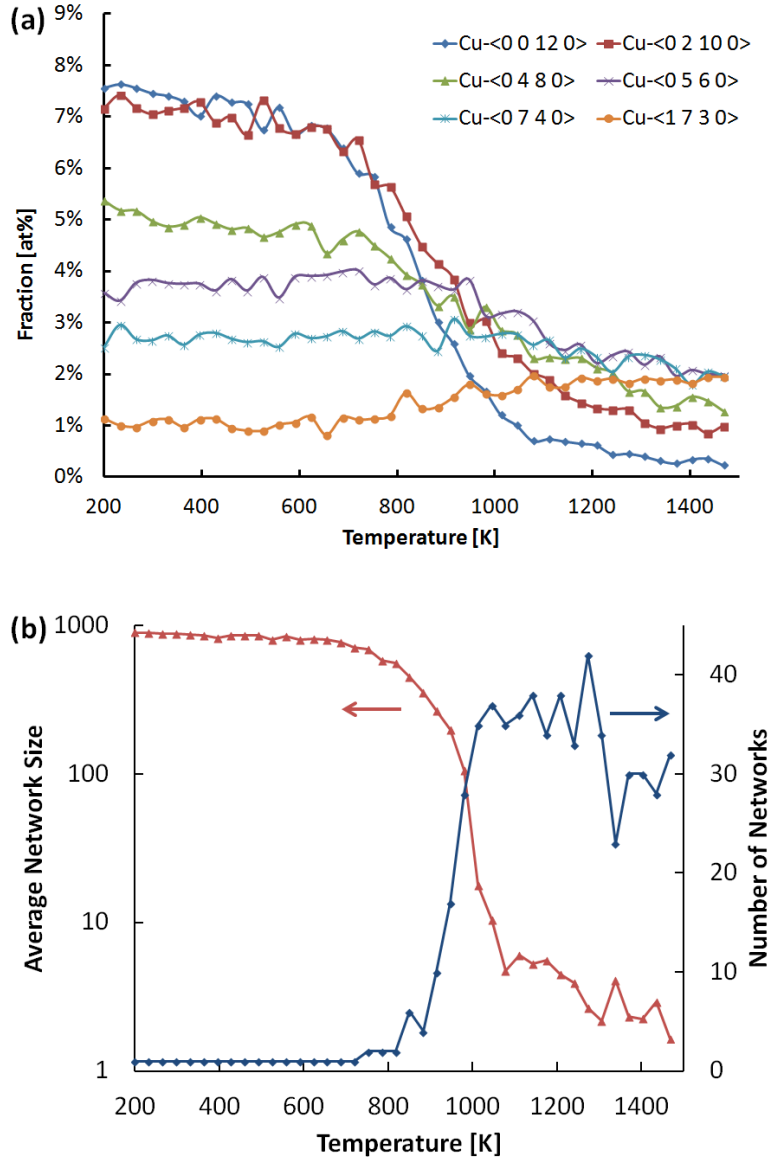

**Figure 3.** (a) The most prevalent clusters by Voronoi index and (b) average size and number of networks of icosahedral clusters in a $Cu_{64.3}Zr_{35.7}$ metallic glass as a function of temperature during a quench at $10^{11}$ K/s.

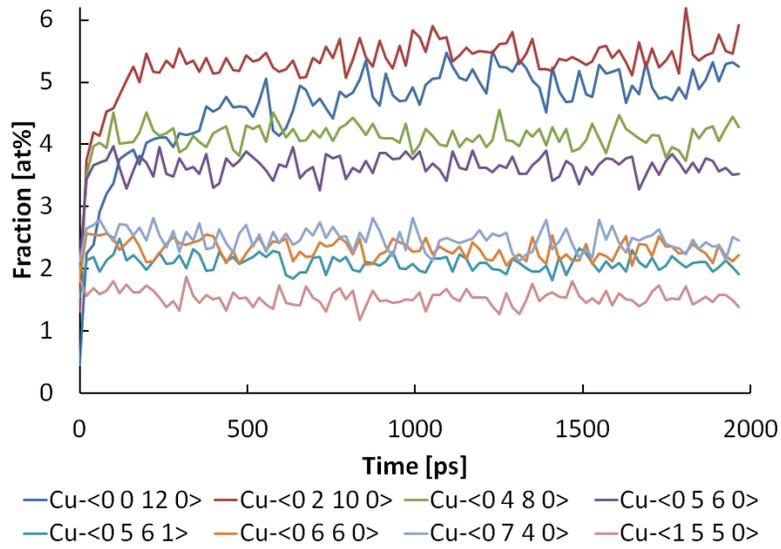

**Figure 4.** Fraction of top eight most prevalent Cu-centered clusters in $Cu_{64.3}Zr_{35.7}$ after a rapid quench from 1500 K to 800 K as a function of time during a constant volume and temperature hold at 800 K.

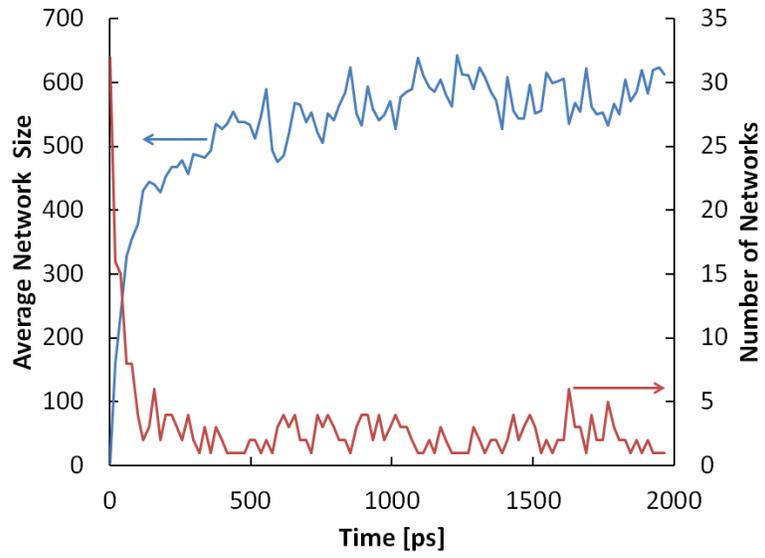

**Figure 5.** Evolution of networks of icosahedral clusters in a $Cu_{64.3}Zr_{35.7}$ metallic glass at 800 K after rapid quench from 1500 K.

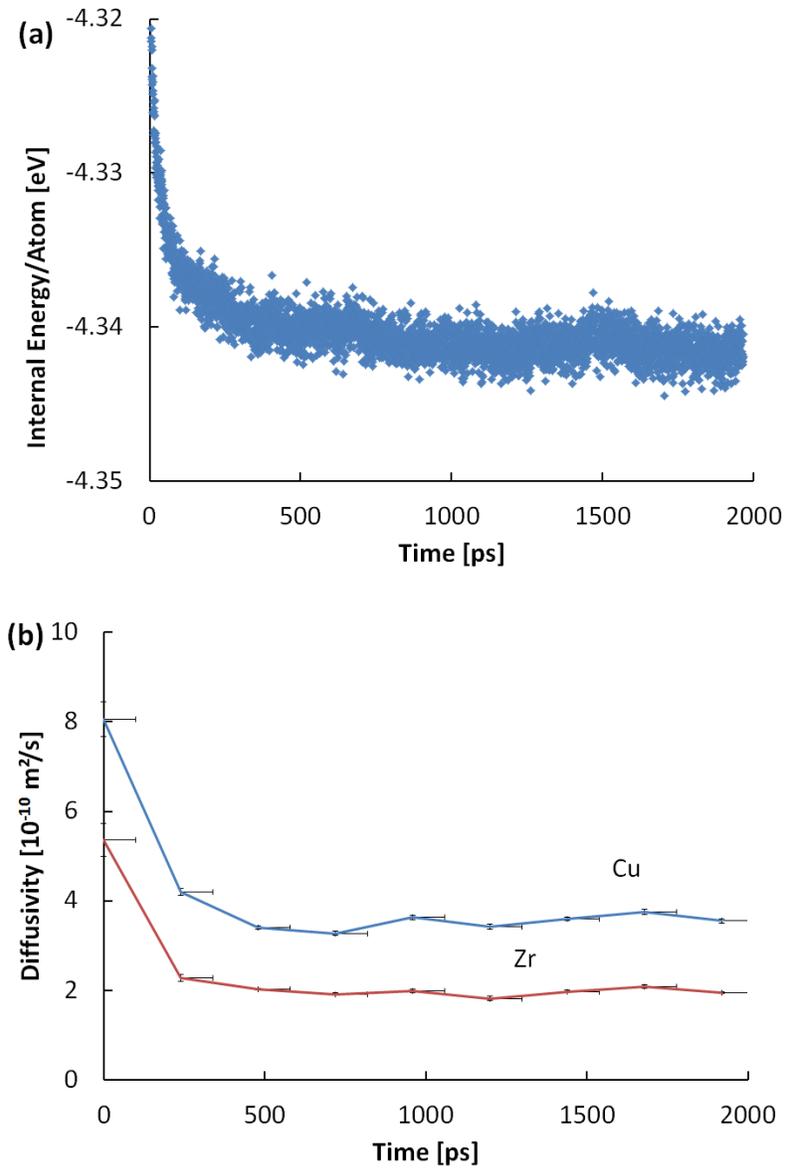

**Figure 6.** (a) Energy per atom and (b) diffusivity in $Cu_{64.3}Zr_{35.7}$ after a rapid quench from 1500 K to 800 K as a function of time during a constant volume and temperature hold at 800 K. Error bars represent the time span used to calculate the diffusivity, and the confidence interval from the diffusivity calculation.

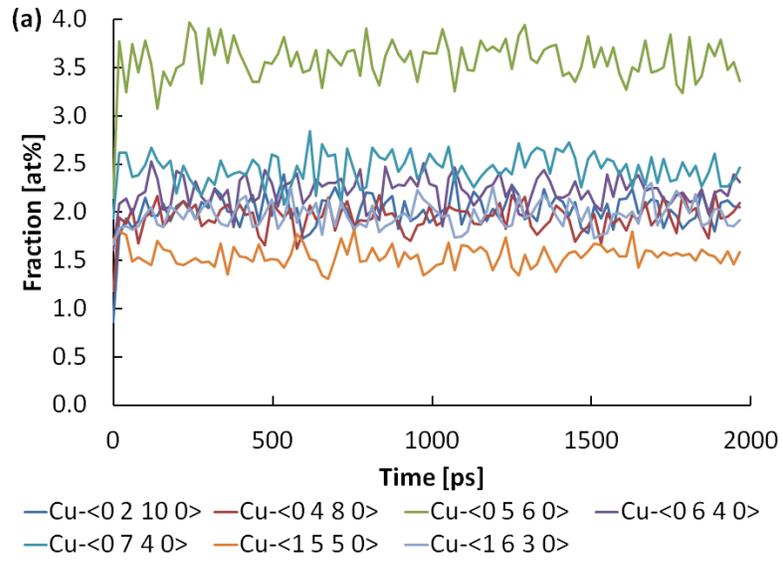

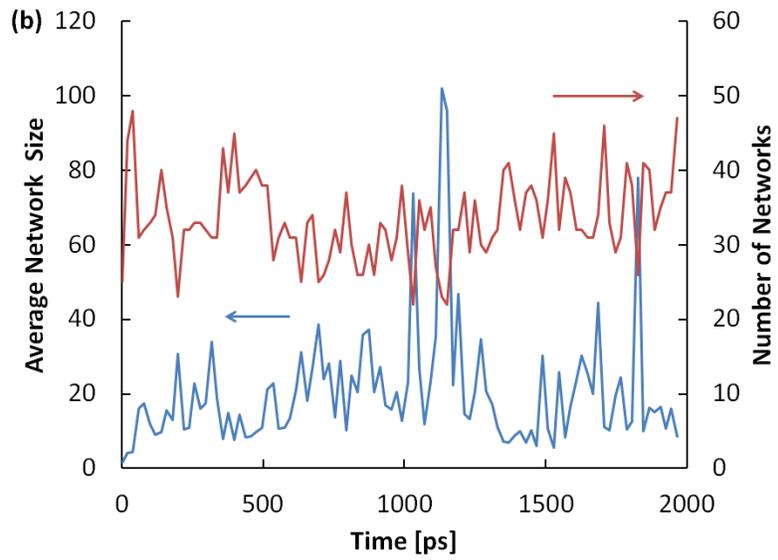

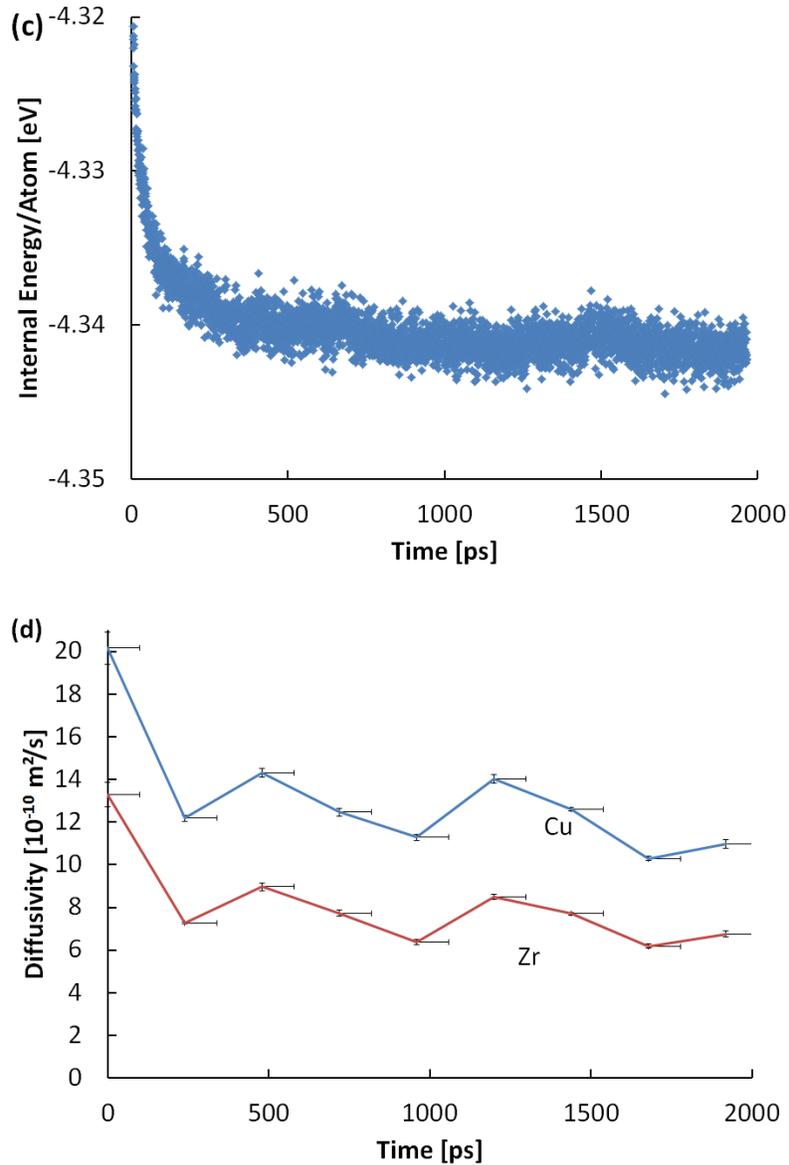

**Figure 7.** Changes in (a) five most prevalent cluster topologies and icosahedral clusters, (b) networks of icosahedral clusters, (c) energy per atom, and (d) diffusivity as a function of time for a $Cu_{50}Zr_{50}$ liquid after rapid quenching from 1500 K to 800 K. Error bars represent the time span used to calculate the diffusivity, and the confidence interval from the diffusivity calculation.

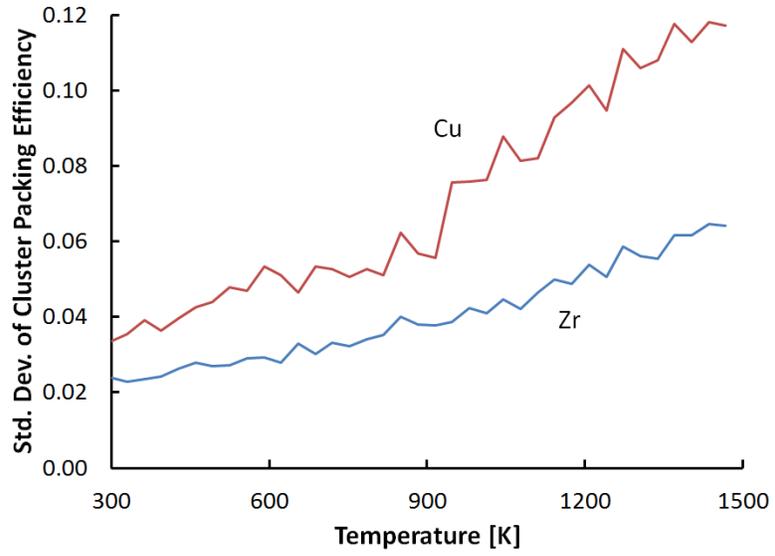

**Figure 8:** Standard deviation of cluster packing efficiency as a function of temperature in a $Cu_{64.3}Zr_{35.7}$ metallic glass during a simulated quench from 1500 K to 300 K at $10^{11}$ K/s.

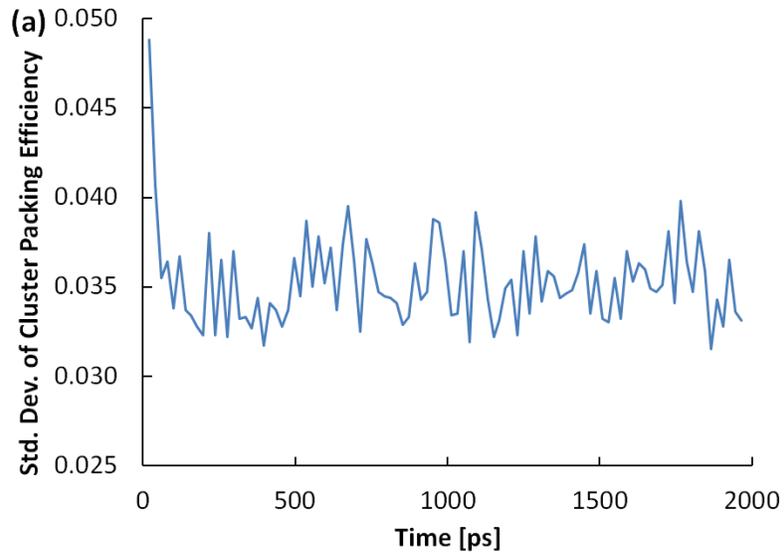

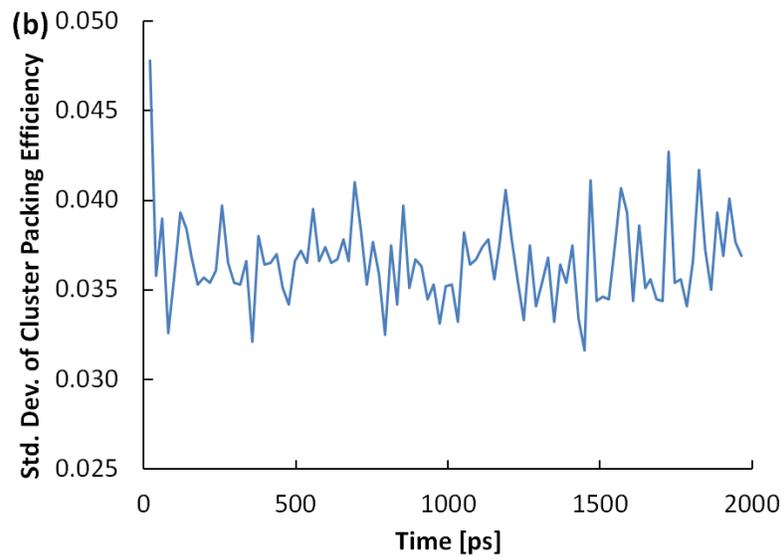

**Figure 9:** Standard deviation of the cluster packing efficiency of Cu-centered clusters for a (a) $Cu_{64.3}Zr_{35.7}$ and (b) $Cu_{50}Zr_{50}$ liquid during a constant volume and temperature hold at 800 K immediately following a rapid quench from 1500 K.

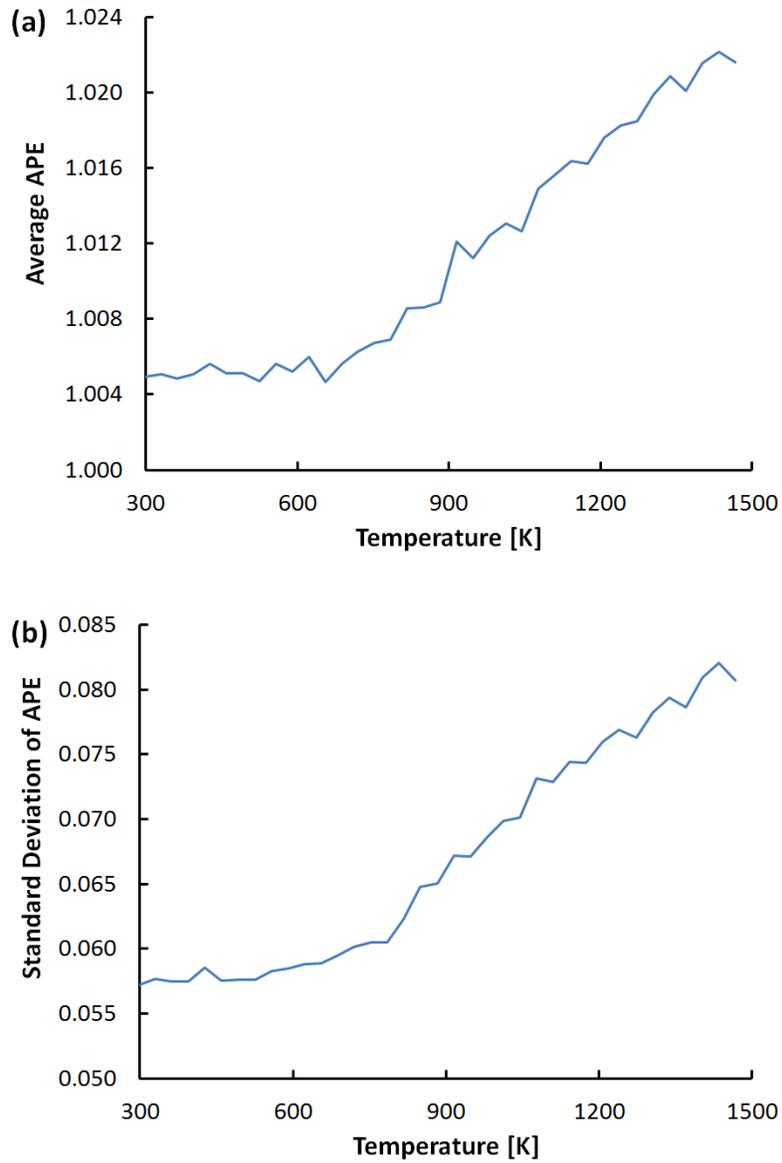

**Figure 10:** (a) Mean and (b) standard deviation of the Atomic Packing Efficiency (APE) of each cluster in a model of a $Cu_{64.3}Zr_{35.7}$ metallic glass during a quench from 1500 K to 300 K at $10^{11}$ K/s.

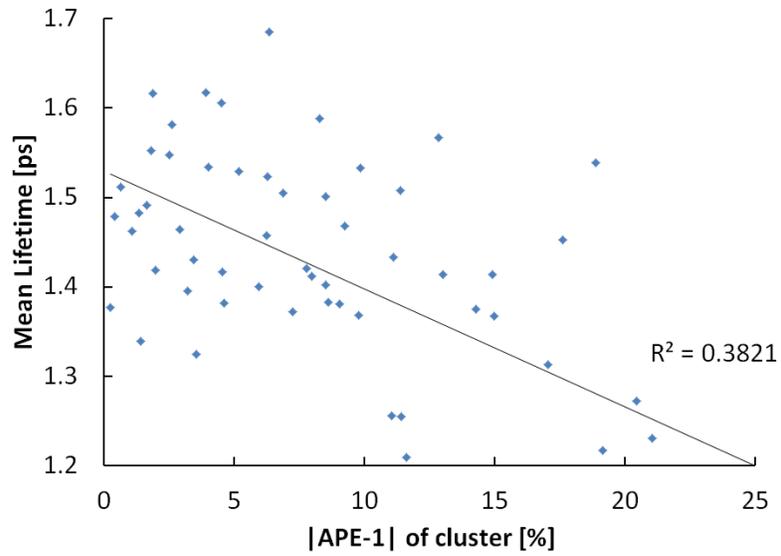

**Figure 11.** Cluster lifetime as a function deviation of APE from ideal in a $Cu_{50}Zr_{50}$ liquid at 800 K. Only clusters that have more than 100 measurements of lifetimes are included.

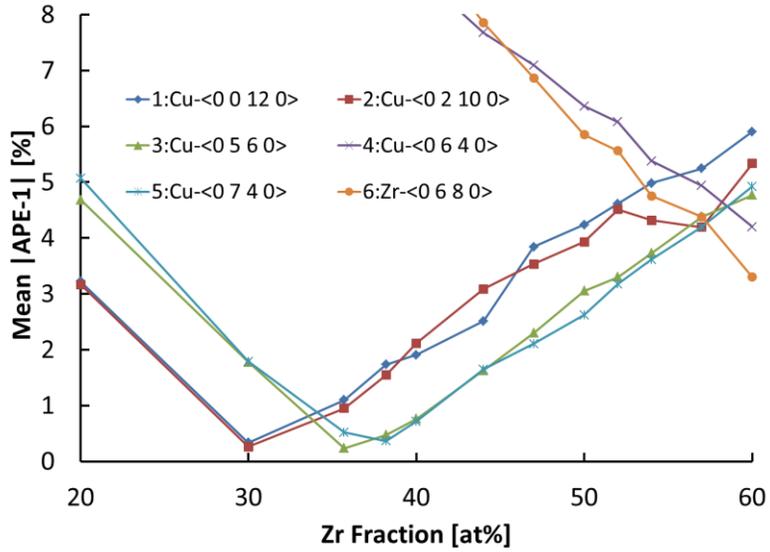

**Figure 12.** Mean deviation from ideal packing efficiency for the six most common cluster types across composition region in Cu-Zr, as measured using the Atomic Packing Efficiency (APE) parameter[18] where an APE of unity is ideally packed.